\documentclass[final,5p,times,twocolumn,authoryear]{elsarticle}

\usepackage{color}
\usepackage{colortbl}
\usepackage{graphics}
\usepackage{amssymb}
\usepackage{amsmath}
\journal{Icarus}

\begin{document}

\begin{frontmatter}



\title{Timescale of asteroid resurfacing by regolith convection resulting from the impact-induced global seismic shaking}


\author{Tomoya M. Yamada\footnote{Corresponding author. Tel.:+81 52 789 3014; fax: +81 52 789 3013. E-mail address: yamada.tomoya@a.mbox.nagoya-u.ac.jp}, Kousuke Ando, Tomokatsu Morota, Hiroaki Katsuragi}

\address{Department of Earth and Environmental Sciences, Nagoya University, Furocho, Chikusa, Nagoya 464-8601, Japan}

\begin{abstract}
A model for the asteroid resurfacing by regolith convection is built to estimate its timescale. In the model, regolith convection is driven by the impact-induced global seismic shaking. The model consists of three steps: (i) intermittent impact of meteoroids, (ii) impact-induced global vibration (seismic shaking), and (iii) vibration-induced regolith convection. In order to assess the feasibility of the resurfacing process driven by regolith convection, we estimate the resurfacing timescale as a function of the size of a target asteroid. In the estimate, a set of parameter values is assumed on the basis of previous works. However, some of them (e.g., seismic quality factor $Q$, seismic efficiency $\eta$, and seismic frequency $f$) are very uncertain. Although these parameter values might depend on asteroid size, we employ the standard values to estimate the representative behavior. To clarify the parameter dependences, we develop an approximated scaling form for the resurfacing timescale. According to the estimated result, we find that the regolith-convection-based resurfacing timescale is shorter than the mean collisional lifetime in most of the parameter uncertainty ranges. These parameter ranges are within those reported by previous works for small asteroids. This means that the regolith convection can be a possible mechanism for the asteroid resurfacing process. 
\end{abstract}


\end{frontmatter}


\section{Introduction}
\label{sec:introduction}
The history of surface evolution of small asteroids is recorded on their surface terrain. Impacts of various-sized meteoroids are the most important processes to discuss the asteroid surface evolution. For instance, regolith grains are generated and mixed by repeated impact events. This {\it gardening} effect keeps the surface fresh and active (e.g.,~\citet{Melosh2011}). 

However, recent planetary explorations have revealed a rich variety of surface terrains that cannot be understood only by the impact cratering and resultant gardening effect. For instance, asteroid Eros, which was explored by NASA's space probe {\it NEAR Shoemaker} from February 2000, showed the interesting characteristics. Based on the observational result, \citet{Asphaug2001} reported the surface blocks distribution that cannot be explained by the potential source craters. To explain the distribution, they discussed the possibility of the size sorting in the vibrated regolith bed. More recently, \citet{Miyamoto2007} and \citet{Tancredi2015} also reported that the surface regolith of asteroid Itokawa, which was explored by the Japanese space probe {\it Hayabusa}, could migrate and be segregated. In addition, impact craters on the surface of Itokawa are very vague~\citep{Saito2006, Hirata2009}. The analysis of tiny samples returned from Itokawa also supports regolith migration. \citet{Tsuchiyama2011} observed some round-shape particles that might be worn away by friction between regolith grains. Moreover, the cosmic-ray exposure (CRE) age of Itokawa's returned particles are measured by~\citet{Nagao2011} and~\citet{Meier2014}. According to these studies, the upper limit of CRE age is approximately $8$ Myr and its probable value is 1.5 Myr. This value is relatively young compared to the CRE age of most of the LL chondrites. All the above-mentioned observations and findings suggest that the regolith migration process plays a crucial role in the surface evolution on small asteroids such as Eros and Itokawa.

For the driving force of regolith migration on Itokawa,~\citet{Miyamoto2007} followed the idea of \citet{Asphaug2001}; they considered the size segregation of vibrated regolith layer. They supposed that the impact-induced global seismic shaking could result in the convection of regolith grains. Moreover, the size segregation of regolith grains could also be driven by regolith convection. Although the impact-induced seismic shaking has not been directly observed on asteroids,~\citet{Richardson2004} and~\citet{Richardson2005} developed a model of the global seismic shaking on asteroid Eros to investigate the crater erasure process. Recently, \citet{Garcia2015} has also computed seismic wave propagation on asteroids in kilometric and sub-kilometric size ranges. Using the global seismic shaking model of~\citet{Richardson2004} and~\citet{Richardson2005},~\citet{Miyamoto2007} showed that the regolith convection induced by global seismic shaking can be caused even by small-scale impacts. For instance, 1 cm impactor with the impact velocity $v_{\rm i}=4$ $\rm km~s^{-1}$ is sufficient to achieve the global seismic shaking on Itokawa because the Itokawa's surface gravity is extremely low, $10^{-4}$~m~s$^{-2}$~\citep{Abe2006,Fujiwara2006}. Thus, the effect of collective granular motion on the surface of asteroid could be very important for properly considering the surface evolution~\citep{Asphaug2007,Katsuragi2016}.

In fact, granular convection can readily be observed when the granular matter is subjected to the vertical vibration in the laboratory experiment (e.g.,~\citet{Yamada2014}). In granular convection induced by vertical vibration, the most important parameter is a ratio between the maximum vibrational acceleration and gravitational acceleration, $\Gamma=A_{0}(2\pi f)^{2}/g$, where, $A_{0}$, $f$ and $g$ are the amplitude of vibration, its frequency, and gravitational acceleration, respectively. The onset criterion of granular convection can be approximately written as $\Gamma > 1$ (e.g.,~\citet{Garcimartin2002}). In addition, if the vibrated granular matter consists of polydisperse grains, the vertical size segregation of grains can also be observed. This vibration-induced size segregation is called Brazil nut effect (BNE). Granular convection is considered a plausible reason for the BNE in the vertically vibrated granular bed~\citep{Knight1993}. 

Recently, the gravity dependence of granular convective motion has been investigated on the basis of laboratory experiments~\citep{Murdoch2013a, Murdoch2013b, Murdoch2013c, Guttler2013, Yamada2014} and numerical simulations~\citep{Tancredi2012,Matsumura2014}. \citet{Murdoch2013a, Murdoch2013b, Murdoch2013c} investigated the gravity dependence of the velocity of shear-driven convective motion by using the parabolic flight method. They reported that the convective velocity is significantly reduced under the microgravity condition while it is enhanced under the high-gravity condition.~\citet{Guttler2013} also performed an experiment of convection-driven BNE using the vibrated granular bed on parabolic flights. They found that the rise velocity of an intruder in the vibrated granular bed is roughly proportional to the gravitational acceleration $g$. \citet{Tancredi2012} numerically computed the threshold velocity $v_{\rm th}$ to induce BNE and reported $v_{\rm th} \sim g^{0.42}$. It should be noticed that, however, $v_{\rm th}$ is the threshold velocity above which segregation starts.~\citet{Matsumura2014} numerically investigated the convection-driven BNE and found that the rise velocity of an intruder in the vibrated granular bed is proportional to $g^{0.5}$. In order to evaluate the gravity dependence of granular convective velocity, the scaling method has been used in~\citet{Yamada2014}. The researchers investigated the convective velocity under the steady vertical vibration on the ground. Using an experimentally obtained scaling relation, they found that the convective velocity is scaled by $g^{0.97}$. While the intruder's rise velocity (BNE rate) and convective velocity are different quantities, \citet{Yamada2014} is roughly consistent with~\citet{Guttler2013}. However, the reason of difference between \citet{Guttler2013} and \citet{Matsumura2014} remains unclear. In any case, all these experimental and numerical results suggest that the granular convection can occur with very low velocity even under the microgravity condition.

In general, the characteristic timescale for gravity-related phenomena is elongated under the microgravity condition. For instance, the penetration timescale of an intruder into a granular target increases under the low-gravity condition~\citep{Altshuler2014}. For granular (regolith) convection, the characteristic timescale would become very long under the microgravity condition because the convective velocity, which is inversely proportional to the timescale, is significantly reduced under the microgravity condition as mentioned above. Therefore, it is not obvious whether the low convective velocity can really resurface the various-sized S-type asteroids covered with regolith within the reasonable timescale. The estimate of timescale is necessary to assess the feasibility of the resurfacing by regolith convection. The main purpose of this study is to build a model of resurfacing process induced by regolith convection for asteroids covered with regolith. Using the model, timescale of the resurfacing can be estimated. Then, the resurfacing timescale should be compared with other timescales such as the mean collisional lifetime. Based on the comparison, we discuss the feasibility of the resurfacing process driven by regolith convection on the surface of asteroids. 

This paper is composed of following sections. In the next section, we build a model of the resurfacing process. Section~\ref{sec:result} presents the estimate of resurfacing timescale. In addition, the scaling analysis for the estimated timescale is discussed in this section. In section~\ref{sec:discussion}, we compare the resurfacing timescale with other timescales and verify the feasibility of the regolith-convection-driven resurfacing. Section~\ref{sec:conclusions} mentions conclusions.

\section{Model}
\label{sec:model}
\subsection{Outline of the resurfacing model}
\label{subsec:outline}
\begin{figure}
\begin{center}
\includegraphics[width=\hsize]{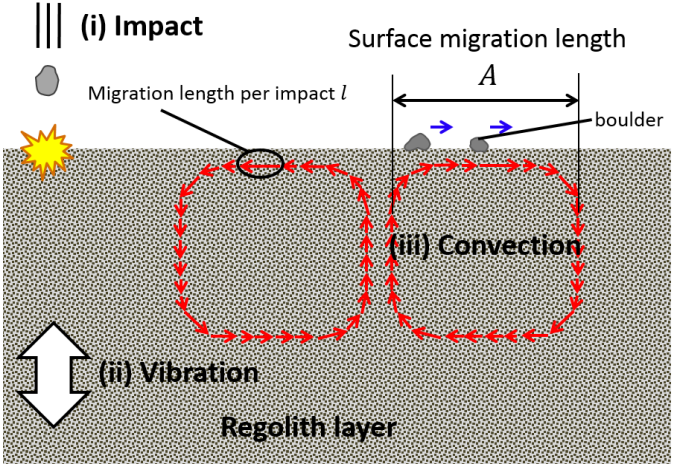}
\end{center}
\caption{Schematic illustration of the model for the asteroid resurfacing induced by regolith convection. Intermittent regolith convection results in the resurfacing of asteroids through three steps: (i) impact, (ii) vibration, and (iii) convection. Each red arrow's length conceptually shows the convective migration length per impact event, $l$. We assume that the convective roll size $A$ corresponds to the surface migration length of the convecting regolith.}
\label{fig:resurface}
\end{figure}

In order to estimate the resurfacing timescale, we make a simple model of the asteroid resurfacing process driven by regolith convection. The regolith convection is caused by impact-induced global seismic shaking. In the model, we divide the resurfacing process into three steps as follows:
\begin{enumerate}
\renewcommand{\labelenumi}{(\roman{enumi})}
\item Impact: Impactors intermittently collide with a target asteroid,
\item Vibration: Each collision leads to a seismic shaking,
\item Convection: The seismic shaking induces the regolith convection on the asteroid.
\end{enumerate}
Fig.~\ref{fig:resurface} shows the schematic illustration of the model. Repeating the series of three steps, the asteroids can be resurfaced by regolith convection. 

In the convection step, we assume that the iteration of small convective migrations constitutes the well-ordered (reproducible) convective roll structure. In Fig.~\ref{fig:resurface}, each red arrow corresponds to the migration length driven by each impact event, and they collectively construct the well-ordered identical convective roll. Although one may feel that this assumption is not so plausible, some experimental works have reported that the intermittently tapped granular bed can construct the well-ordered convective motion or BNE~\citep{Philippe2003,Philippe2005,Tancredi2012,Iikawa2015}.

The details of steps (i), (ii), and (iii) are written in sections~\ref{subsubsec:impactor},~\ref{subsubsec:shaking}, and~\ref{subsubsec:convection}, respectively. Finally, we unify these three steps and compute the resurfacing timescale. The unification is described in section~\ref{subsec:timescale}.  

\subsubsection{Impact frequency}
\label{subsubsec:impactor}
\begin{figure}
\begin{center}
\includegraphics[width=\hsize]{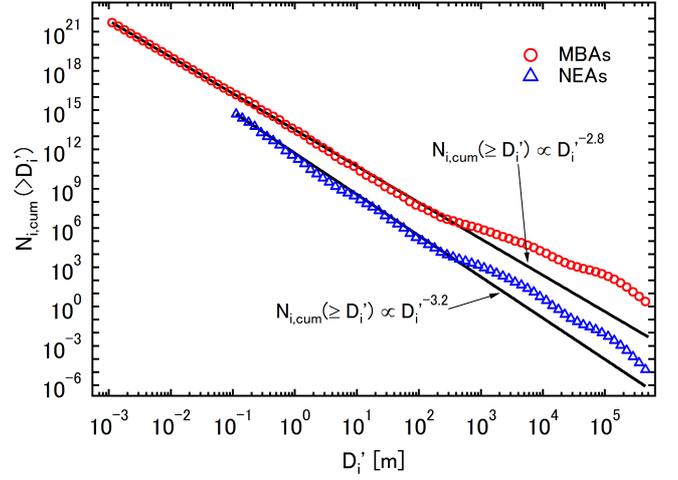}
\end{center}
\caption{Cumulative number distribution $N_{\rm i,cum}(\geq D_{\rm i}')$ vs.~impactor diameter $D_{\rm i}'$. Red open circles and blue open triangles respectively show the numerical data for MBAs and NEAs computed by~\citet{O'Brien2005}. The solid lines are power-law fits.} 
\label{fig:Nicum_Di}
\end{figure}

The frequency of impacts per year $N_{\rm p}$ is a necessary quantity to build the model. $N_{\rm p}$ is estimated from the population of small objects $N_{\rm i}$ of diameter $D_{\rm i}$. To estimate $N_{\rm i}$ for the Main Belt Asteroid (MBA) and Near-Earth Asteroid (NEA), there have been two models based on the dynamical evolution of the main belt population:~\citet{O'Brien2005} (called hereafter OBG population) and \citet{BottkeJr2005} (called hereafter BAL population). The range of estimated impactor size in OBG population is broader than that of BAL population, and both are in good agreement with the currently observed population in the main belt. Only the minor differences can be recognized between these two models. Here, we employ the OBG population to compute $N_{\rm p}$. Fig.~\ref{fig:Nicum_Di} shows the cumulative number distribution of potential impactors $N_{\rm i,cum}(\geq D'_{\rm i})$ for MBAs and NEAs, estimated by OBG population.

By numerically differentiating $N_{\rm i,cum}$, we can obtain the incremental distribution of $N_{\rm i}$ of small bodies with diameters between ${D'_{{\rm i}, k}}$ and ${D'_{{\rm i}, k+1}}$, where $k$ is the index of the binned data (${D'_{{\rm i}, k+1}}>{D'_{{\rm i}, k}}$). The relation between $N_{\rm i,cum}(\geq{D'_{{\rm i}, k}})$ and $N_{\rm i}(D_{{\rm i}, k})$ is described as
\begin{equation}
N_{\rm i}(D_{{\rm i}, k})=N_{\rm i,cum}(\geq{D'_{{\rm i}, k}}) -N_{\rm i,cum}(\geq{D'_{{\rm i}, k+1}}).
\end{equation}
Here, we consider the representative value of size $D_{{\rm i}, k}=\sqrt{{D'_{{\rm i}, k+1}}{D'_{{\rm i}, k}}}$ because the OBG population data adapted logarithmic bin size for ${D'_{{\rm i}, k}}$.

By multiplying $N_{\rm i}$ by the annual mean collision probability per collisional cross-section $P_{\rm i}$ and the collisional cross-section $(D_{\rm i}/2 + D_{\rm a}/2 )^{2}$, we can compute the number of impacts per year $N_{\rm p}$. The specific form of $N_{\rm p}$ is written as
\begin{equation}
N_{\rm p}(D_{\rm i}, D_{\rm a})=N_{\rm i}(D_{\rm i})P_{\rm i}\left( \frac{D_{\rm i}}{2} +\frac{D_{\rm a}}{2} \right)^{2},
\label{eq:I_frequency}
\end{equation}
where $D_{\rm a}$ is the diameter of target asteroid. Since this procedure is applied to all $k$ data, we omit the subscript $k$ hereafter. In this study, we use the constant values $P_{\rm i}=2.9 \times 10^{-24}$ ${\rm m^{-2}yr^{-1}}$ for the collision of MBAs~\citep{Bottke1993,Bottke1994b} and $P_{\rm i}=15 \times 10^{-24}$ ${\rm m^{-2}yr^{-1}}$ for the collision of NEAs~\citep{Bottke1994a}. Thus, $N_{\rm p}$ can merely be defined as a function of $D_{\rm i}$ and $D_{\rm a}$~\citep{O'Brien2005}. Actually, the value of $P_{\rm i}$ depends on the location and orbit of small bodies. Here we neglect such higher-order effects and consider the mean value.

\subsubsection{Global seismic shaking}
\label{subsubsec:shaking}
Next, the strength of seismic shaking $\Gamma$, which corresponds to the maximum vibration acceleration normalized to the gravitational acceleration, has to be estimated for each impact event. Based on previous reports, we assume that the regolith convection can be induced by the global seismic shaking satisfying $\Gamma>1$. Moreover, we use the previous experimental result, in which the granular convective velocity is scaled by $\Gamma$~\citep{Yamada2014}, to compute the regolith convective velocity. 

\begin{figure}
\begin{center}
\includegraphics[width=\hsize]{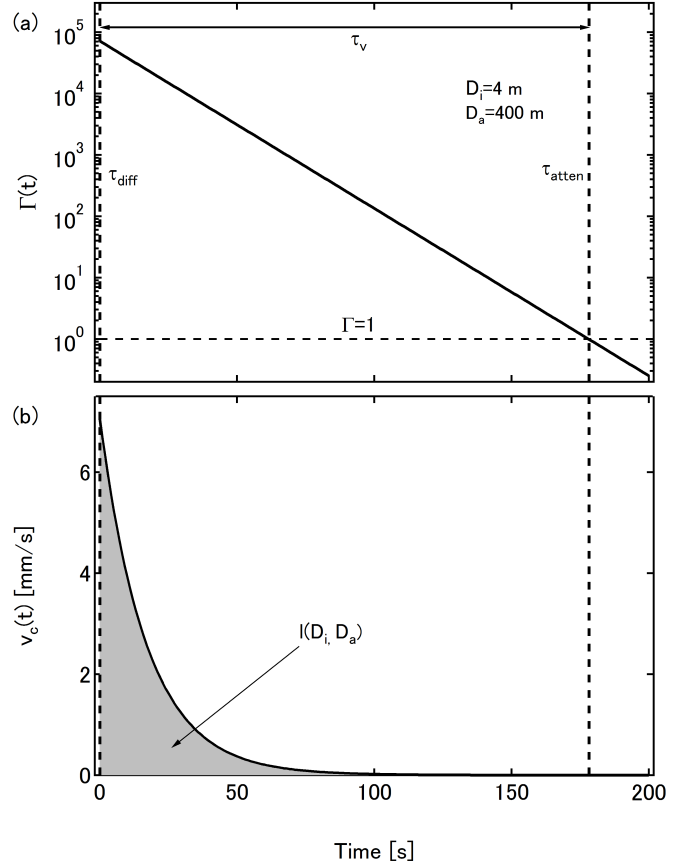}
\end{center}
\caption{
(a) Temporal decay of the maximum dimensionless acceleration $\Gamma$ after the impact. The solid line shows an example of the case of $D_{\rm i}=4$ m, $D_{\rm a}=400$ m, and $v_{\rm i}=5.3$, $\rm km~s^{-1}$. To compute $\Gamma(t)$, we use the standard parameter values: $Q=2000$, $\eta=10^{-4}$, $f=20$ Hz, $\rho_{\rm i}=1900$ $\rm kg~m^{-3}$, and $\rho_{\rm a}=2500$ $\rm kg~m^{-3}$. The horizontal dashed line corresponds to $\Gamma=1$. Vertical broken lines show $\tau_{\rm diff}$ and $\tau_{\rm atten}$, and a two-way arrow represents $\tau_{\rm v}$ defined by Eq.~(\ref{eq:v}). (b) Temporal decay of the convective velocity $v_{\rm c}$. The solid curve is computed by Eqs.~(\ref{eq:accel_Q}) and (\ref{eq:Vc}) with the standard parameter values (Table~\ref{table:parameters1}). The hatched area corresponds to the migration length per impact $l=0.13$ m. Its specific form is written in Eq.~(\ref{eq:l2}). 
}
\label{fig:gamma_t}
\end{figure}

To estimate $\Gamma$ caused by meteoroid impacts, we employ the global seismic shaking model proposed by~\citet{Richardson2004} and~\citet{Richardson2005}.
The model is based on two assumptions. First, the kinetic energy of the impactor is transferred to the seismic energy with the impact seismic efficiency factor $\eta$. Second, the transferred seismic energy is equally distributed to the entire target asteroid by the diffusion of seismic energy density. Namely, the homogeneous seismic energy distribution is considered as an average picture. The distributed seismic energy density attenuates due to the inelasticity of the internal structure of asteroids. The attenuation of seismic energy density is characterized by the seismic quality factor $Q$. In addition, spherical shape of the target asteroid is assumed for the sake of simplicity. Although the numerical and observational evidences for the validity of this model are still very limited, this model offers an analytic way to compute the global seismic shaking on small bodies, and the analytic result was compared with numerical simulations~\citep{Garcia2015}. From these considerations, $\Gamma$ can be calculated as (\citet{Richardson2005}, \citet{Miyamoto2007} and its SOM), 

\begin{equation}
\Gamma (t, D_{\rm i}, D_{\rm a}) = \frac{3fv_{\rm i}}{G}\sqrt{\eta \frac{\rho_{\rm i} D_{\rm i}^{3}}{\rho_{\rm a}^{3} D_{\rm a}^{5}}} {\rm exp} \left(-\frac{2\pi f}{Q}t \right)
\label{eq:accel_Q}
\end{equation}
where $\rho_{\rm a}$ is the density of the target asteroid, $\rho_{\rm i}$ is the density of impactor, $f$ is the dominant seismic frequency, $v_{\rm i}$ is the mean impact velocity, $t$ is the elapsed time from the moment of impact, and $G$ is the gravitational constant. Fig.~\ref{fig:gamma_t}(a) presents an example of $\Gamma(t)$ for $D_{\rm a}=400$ m (Itokawa-sized asteroid), $D_{\rm i}=4$ m, and $v_{\rm i}=5.3$ $\rm km~s^{-1}$~\citep{Bottke1994b, O'Brien2005}. Here, the following parameter values are also employed: $\rho_{\rm a}=1900$ $\rm kg~m^{-3}$, $\rho_{\rm i}=2500$ $\rm kg~m^{-3}$, $Q=2000$, $\eta=10^{-4}$ and $f=20$ Hz~\citep{Richardson2004, Richardson2005, Fujiwara2006, Miyamoto2007, Yasui2015}. Hereafter, we basically use these parameter values as standard ones in the following computations unless otherwise noted. Because we mainly focus on the porous rubble-pile or very fractured target asteroids, the relatively low density for the target $\rho_{\rm a}=1900$ $\rm kg~m^{-3}$ is utilized as the standard one. For the dominant seismic frequency,~\citet{Richardson2004} and \citet{Richardson2005} mentioned that it falls in $f=10 - 20$ Hz for asteroid Eros due to the fractured inner structures which act as natural band-pass filter for seismic frequency. In addition, they assumed $Q=2000$ on the asteroid Eros based on the similarity between internal structure of fractured S-type asteroids and upper lunar crust. Recently,~\citet{Garcia2015} has represented that the dominant frequency should increase as the asteroid size decreases. In addition, they reported that the relatively high frequency ($\sim 100$ Hz) dominates the vibration on Itokawa-sized asteroid when the monolithic core structure and low dissipation ($Q=2000$) are assumed. The reason for this discrepancy is the difference in assumption of the asteroid internal structure. Moreover, its value could vary depending on inhomogeneity of the internal structure. To consider the diffusion of seismic energy density, the highly inhomogeneous structure must be assumed. Here, we temporarily employ the constant value as standard ones mainly on the basis of~\citet{Richardson2004, Richardson2005}. However, the details of parameter dependences will be discussed in section~\ref{subsubsec:scaled_parameter}. The standard parameter values are summarized in Table~\ref{table:parameters1}.

\begin{table*}[hbtp]
\caption{Standard parameter values for the global seismic shaking of small bodies. Mean impact velocity: $v_{\rm i}$, seismic quality factor: $Q$, impact seismic efficiency: $\eta$, dominant seismic frequency: $f$, density of the target asteroid: $\rho_{\rm a}$, density of impactor: $\rho_{\rm i}$, and diffusion constant of seismic energy density: $K_s$.}
  \label{table:parameters1}
  \small
  \centering
  \begin{tabular}{cccccccc}
    \hline
  $v_{\rm i}$ ($\rm kms^{-1}$) & $Q$ & $\eta$ & $f$ (Hz) & $\rho_{\rm i}$ ($\rm kg~m^{-3}$) & $\rho_{\rm a}$ ($\rm kg~m^{-3}$) & $K_s$ (m$^2$~s$^{-1}$) \\
    \hline \hline
	5.3 & 2000 & $10^{-4}$ & 20 & 1900 & 2500 & $2.5 \times 10^5$ \\
    	\hline
  \end{tabular}
\end{table*}

Then, we compute the duration of global seismic shaking $\tau_{\rm v}$ which is defined by a period of $\Gamma > 1$ in the entire asteroid. The specific form of $\tau_{\rm v}$ is written as, 
\begin{equation}
\tau_{\rm v}(D_{\rm i}, D_{\rm a})=\tau_{\rm atten}(D_{\rm i}, D_{\rm a}) - \tau_{\rm diff}(D_{a}),
\label{eq:v}
\end{equation}
where $\tau_{\rm atten}(D_{\rm i}, D_{\rm a})$ is the period from the moment of impact to the time at which $\Gamma$ decays to be $1$ due to the inelastic dissipation. And, $\tau_{\rm diff}(D_{\rm a})$ represents the time required to diffuse the seismic energy density to the entire asteroid. 

\begin{figure}
\begin{center}
\includegraphics[width=\hsize]{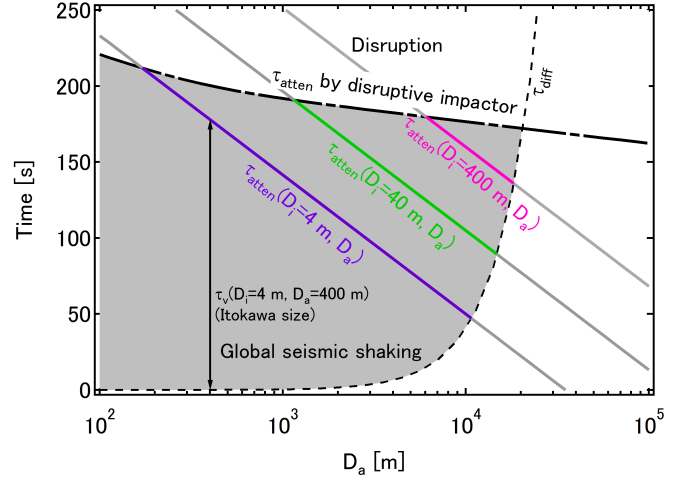}
\end{center}
\caption{The relationships among various characteristic times for MBAs. The solid lines indicate the attenuating duration $\tau_{\rm atten}$ by the impactor with diameter $D_{\rm i}=$ $400$ m (pink), $40$ m (green), and $4$ m (purple) vs.~the target asteroid diameter $D_{\rm a}$. The parameter values used are identical to those in Fig.~\ref{fig:gamma_t}. The dashed curve shows $\tau_{\rm diff}$ with $K_{s}=0.25$ km$^{2}$~s$^{-1}$~\citep{Richardson2004, Richardson2005, Miyamoto2007}. The chain curve shows $\tau_{\rm atten}$ by the largest impactor which disrupts itself~\citep{Jutzi2010}. The global seismic shaking is attained in the hatched area. A two-way arrow shows the vibrated duration $\tau_{\rm v}=\tau_{\rm atten} - \tau_{\rm diff}$ for the collision between the impactor of $D_{\rm i}=$ $4$ m and target of $D_{\rm a}=$ $400$ m (Itokawa size).}
\label{fig:tv_Da}
\end{figure}

The form of $\tau_{\rm atten}$ can be computed from Eq.~(\ref{eq:accel_Q}) as, 
\begin{equation}
\tau_{\rm atten}(D_{\rm i}, D_{\rm a})=\frac{Q}{2\pi f} {\ln} \left( \frac{3fv_{\rm i}}{G}\sqrt{\eta \frac{\rho_{\rm i}D_{\rm i}^{3}}{\rho_{\rm a}^{3}D_{\rm a}^{5}}} \right).
\label{eq:atten}
\end{equation}
The value of $\tau_{\rm atten}$ depends on both $D_{\rm i}$ and $D_{\rm a}$. In Fig.~\ref{fig:tv_Da}, $\tau_{\rm atten}(D_{\rm a})$ with $D_{\rm i}=400$ m (pink solid line), $D_{\rm i}=40$ m (green solid line), and $D_{\rm i}=4$ m (purple solid line) are shown. All other parameter values are fixed to the standard ones in these lines. 

On the other hand, $\tau_{\rm diff}$ is the diffusion timescale of the seismic energy density (\citet{Richardson2004} and~\citet{Richardson2005}). The computed form of $\tau_{\rm diff}$ is written as, 
\begin{equation}
\tau_{\rm diff}(D_{\rm a})=\frac{D_{\rm a}^{2}}{K_{\rm s}\pi^{2}}.
\label{eq:tdff}
\end{equation}
In Fig.~\ref{fig:tv_Da}, $\tau_{\rm diff}(D_{\rm a})$ is represented by the dashed curve. Here, we employ the typical value for the diffusion coefficient $K_{\rm s}=2.5 \times 10^{5}$ m$^{2}$~s$^{-1}$~\citep{Richardson2004,Richardson2005}. The global seismic shaking is achieved in the area above the dashed curve ($\tau_{\rm diff}(D_{\rm a})$) and below the $\tau_{\rm atten}(D_{\rm i}, D_{\rm a})$. For instance, the two-way arrow in Fig.~\ref{fig:tv_Da} shows $\tau_{\rm v}$ for the collision of $D_{\rm i}=4$ m to $D_{\rm a}=400$ m. The corresponding durations $\tau_{\rm atten}$, $\tau_{\rm diff}$ and $\tau_{\rm v}$ are also shown in Fig.~\ref{fig:gamma_t}. 

If $\tau_{\rm atten}$ is smaller than $\tau_{\rm diff}$, the global seismic shaking is impossible. In such a situation, only the {\it local} seismic shaking is achieved in the vicinity of the impact point. While the global seismic shaking is attained at $t=\tau_{\rm diff}$, the local seismic shaking occurs even in the early stage of the seismic shaking, $t<\tau_{\rm diff}$. This is particularly the case for the large target asteroids of $D_{\rm a} > 5 \times 10^3$ m. For the small target asteroids of $D_{\rm a} < 5 \times 10^3$ m, on the other hand, $\tau_{\rm diff}$ is negligibly short as shown in Fig.~\ref{fig:tv_Da}. For example, in the case of Itokawa-sized asteroid, the estimated diffusion timescale $\tau_{\rm diff}(D_{\rm a}=400 \mbox{ m})=6.6 \times 10^{-2}$ s is much shorter than $\tau_{\rm atten}(D_{\rm i}=4 \mbox{ m}, D_{\rm a}=400 \mbox{ m})=180$ s. In this regime, the global seismic shaking is quickly achieved, and $\tau_{\rm v}$ can be simply approximated by $\tau_{\rm atten}$. According to Fig.~\ref{fig:tv_Da}, the asteroids in the range of $5 \times 10^{3}$ m $< D_{\rm a} < 2 \times 10^4$ m can be vibrated globally only for relatively a short time. Moreover, the asteroids in the range of $D_{\rm a} > 2 \times 10^{4}$ m cannot be vibrated globally because $\tau_{\rm diff}$ becomes longer than $\tau_{\rm atten}$ by the largest impactor which disrupts the target. (The chain line in Fig.~\ref{fig:tv_Da}; the definition of disruption limit is explained later in the next subsection.) Even for such large asteroids, the local seismic shaking could be induced at the neighborhood of the impact point.

\subsubsection{Regolith convection}
\label{subsubsec:convection}
Substituting Eq.~(\ref{eq:accel_Q}) into the velocity scaling of the granular convection (Eq.~(8) in \citet{Yamada2014}), we can estimate the velocity of regolith convection. Using the convective velocity scaling and $\tau_{\rm v}$, the migration length per impact $l$ can be computed. Schematic image of $l$ is represented by red arrows in Fig.~\ref{fig:resurface}.

The migration length per impact $l$ can be estimated as follows. The granular convective velocity $v_{\rm c}$ has been characterized as a function of $\Gamma$ in the previous study~\citep{Yamada2014}. The form of $v_{\rm c}(t, D_{\rm i}, D_{\rm a})$ is written as, 
\begin{equation}
v_{\rm c}(t, D_{\rm i}, D_{\rm a})=C_{\rm 0}\left(\frac{\Gamma (t, D_{\rm i}, D_{\rm a})}{2\pi f} \right)^{2\alpha}g^{\alpha +\frac{1}{2}}d^{-\alpha+\frac{1}{2}}\left( \frac{A}{d}\right)^{\beta},
\label{eq:Vc}
\end{equation}
where $d$ is the diameter of grains and $A$ is the convective roll scale. The roll size $A$ is defined by the geometric mean of the horizontal and vertical dimensions $A=\sqrt{RH}$, where $R$ and $H$ are the radius and height of the vibrated granular bed. \citet{Yamada2014} determined the parameter values by the fitting to the experimental data as $C_{\rm 0}=3.6\times 10^{-3}$, $\alpha=0.47$, and $\beta=0.82$. We can obtain the following form of $l(D_{\rm i}, D_{\rm a})$ by integrating Eq.~(\ref{eq:Vc}) from $\tau_{\rm diff}$ to $\tau_{\rm atten}$: 
\begin{eqnarray}
l(D_{\rm i}, D_{\rm a}) &=& \int_{\tau_{\rm diff}}^{\tau_{\rm atten}}v_{\rm c} \left( \Gamma (t, D_{\rm i}, D_{\rm a}) \right) dt \label{eq:l1} \\
			          &=& \frac{C_{\rm 0}Qg^{\alpha +\frac{1}{2}}d^{-\alpha-\beta+\frac{1}{2}}{A^{\beta}}}{2\alpha(2\pi f)^{1+2\alpha}} \nonumber \\
&& \times  \left[ \left( \frac{3fv_{\rm i}}{G}\sqrt{\eta \frac{\rho_{\rm i} D_{\rm i}^{3}}{\rho_{\rm a}^{3} D_{\rm a}^{5}}} \right)^{2\alpha} {\exp}\left( -\frac{4\alpha \pi f \tau_{\rm diff}}{Q} \right) -1 \right]. 
\label{eq:l2}
\end{eqnarray}
Fig~\ref{fig:gamma_t}(b) shows $v_{\rm c}(t)$ calculated from Eqs.~(\ref{eq:accel_Q}) and (\ref{eq:Vc}). In Fig.~\ref{fig:gamma_t}(b), the hatched area corresponds to $l(D_{\rm i}= 4 {\mbox m}, D_{\rm a}= 400 {\mbox m})=0.13$ m. The values of $C_{\rm 0}$, $\alpha$, and $\beta$ were obtained on the basis of spherical glass-beads experiments performed on the ground. Although it might not be appropriate to use these values for regolith migration, here we employ these values for the first step approximation. Furthermore, parameter sensitivities in the estimate of resurface timescale will be discussed later in section \ref{subsec:feasibility}

To estimate the specific values of $v_{\rm c}$ and $l$ using Eqs.~(\ref{eq:Vc}) and (\ref{eq:l2}), we have to assume the values of $d$ and $A$ in addition to $v_{\rm i}$, $Q$, $\eta$, $f$, $\rho_{\rm i}$, and $\rho_{\rm a}$. We use $1$ cm for the typical value of $d$ based on the actual observation of Itokawa~\citep{Fujiwara2006,Yano2006}. Although $d$ obeys a certain distribution form in general, we use the typical constant value to investigate the representative behavior of the asteroid. For the convective roll size $A$, we assume that $A$ fulfills $A/d=100$. This assumption is based on the experimental observation of granular convection. In \citet{Yamada2014}, a transition from one pair of convective rolls to two pairs of convective rolls was observed when the convective (horizontal) system size is approximately $A/d \simeq 100$. \citet{Aoki1996} also confirmed the splitting of convective rolls at $A/d \simeq 90$. Furthermore, the convective roll most likely has a vertical intrinsic length scale as well. In the deeper part of the vibrated granular bed, the convective velocity becomes almost zero. This quiescent part is called {\it frozen zone}~\citep{Taguchi1992, Ehrichs1995, Knight1996, Yamada2014}. From these experimental observations, we assume that the granular convective roll size satisfies $A/d=100$ both in horizontal and vertical directions (i.e., aspect ratio is 1) in the current model. 

Here, we employ additional restrictions to the range of $D_{\rm i}$ that is relevant to compute the resurfacing timescale. Both the upper and lower limits for $D_{\rm i}$ are introduced by considering the disruption and well-ordered convective motion by intermittent events, respectively.

\begin{figure}
\begin{center}
\includegraphics[width=\hsize]{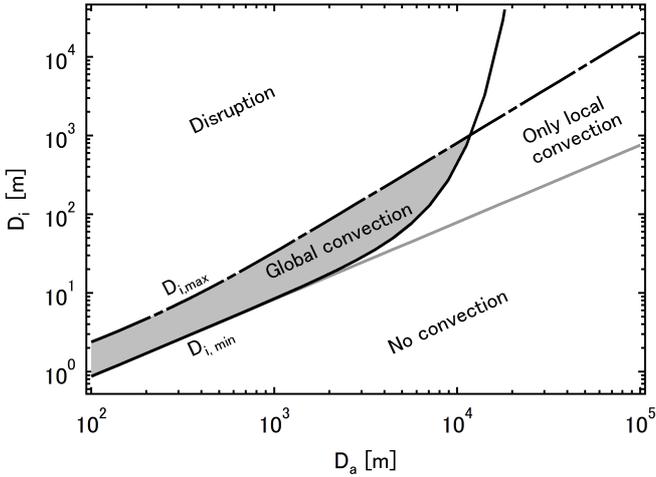}
\end{center}
\caption{A range of impactor size which can generate the regolith convection of MBAs. The chain line represents the upper limit $D_{\rm i, max}$ at which the disruption of the target asteroid occurs~\citep{Jutzi2010}. The solid line represents the lower limit $D_{\rm i, min}$ which is defined by the minimum migration length ($n=0.1$ in Eq.~(\ref{eq:lowerlimit_Di})). The gray line shows $D_{\rm i, min}$ computed with the approximation $\tau_{\rm diff} = 0$.}
\label{fig:Di_range}
\end{figure}

Too large impactors would disrupt the target asteroid. This upper limit value is denoted by $D_{\rm i,max}$. $D_{\rm i, max}$ can be computed from the threshold of catastrophic disruption $Q_{d}^{\ast}$ which corresponds to the critical amount of energy per unit mass delivered by the impactor. The catastrophic disruption means that the target asteroid is pulverized and dispersed until the mass of the largest remnant is smaller than a half of the original mass. $Q_{d}^{\ast}$ for broad-size-ranging basalt targets was self-consistently characterized by \citet{Benz1999} using numerical method. \citet{Jutzi2010} used an improved numerical code, so that $Q_{d}^{\ast}$ for pumice targets, which are not fractured but monolithic with microporosity, can be computed. In addition, it should be noted that all $Q_{d}^{\ast}$ values introduced here are computed with continuum target modeling. In general, it is not easy to estimate the realistic $Q_{d}^{\ast}$ for rubble-pile targets. Since we focus on the porous rubble-pile or very fractured asteroids in this study, the porous-target-based $Q_{d}^{\ast}$ reported by \citet{Jutzi2010} is employed. The obtained $D_{\rm i, max}(D_{\rm a})$ is plotted by the chain curve in Fig.~\ref{fig:Di_range}. $\tau_{\rm atten}$ by the disruptive impactor can also be computed from $D_{\rm i, max}$. The chain curve in Fig.~\ref{fig:tv_Da} indicates this disruption limit $\tau_{\rm atten}(D_{\rm i,max},D_{\rm a})$ above which the target asteroid is pulverized.

On the other hand, too small impactor is not able to trigger the convective motion. In other words, $l$ must be large enough to produce the well-ordered (reproducible) convective motion of the roll size $A$. Thus, we assume that the minimum limit of migration length $l_{\rm min}$ should be related to the convective roll size $A$. If $l_{\rm min}$ is much shorter than $A$, it is probably difficult to follow the same convective structure by a series of intermittent seismic shakings. Therefore, the lower limit $D_{\rm i, min}$ can be written as, 
\begin{equation}
l_{\rm min}=nA,
\label{eq:lowerlimit_Di}
\end{equation}
where $n$ is a parameter satisfying $n>0$. The obtained lower limit $D_{\rm i, min}(D_{\rm a})$ in the case of $n=0.1$ is presented by the solid curve in Fig.~\ref{fig:Di_range}. Consequently, we find that there is a finite area in which the global convective motion can be induced (hatched area in Fig.~\ref{fig:Di_range}). Hereafter, we use $n=0.1$ unless otherwise noted. Since we assume $A/d=100$, $n=0.1$ corresponds to $l_{\rm min}=10d$. Although this $n$ value is speculatively assumed, we evaluate its effect to the estimate of timescale by means of scaling method, later in section~\ref{subsubsec:scaled_parameter}. The gray line in Fig.~\ref{fig:Di_range} corresponds to $D_{\rm i, min}(D_{\rm a})$ computed with the approximation $\tau_{\rm diff} = 0$.

Namely, Fig.~\ref{fig:Di_range} shows the phase diagram of the outcome induced by the impact between $D_{\rm i}$ and $D_{\rm a}$. The hatched area in Fig.~\ref{fig:Di_range} represents the global seismic shaking phase that results in the well-ordered convective motion. Above the chain line ($D_{\rm i, max}$), the target asteroid is destroyed by the impact. Below the solid or gray lines ($D_{\rm i, min}$), the well-ordered convective motion cannot be induced. This limit corresponds to the boundary for the no-convection phase due to the too small impactor. The upper right part bounded by chain, solid, and gray lines indicates the phase in which only the local seismic shaking is achieved. For Itokawa-sized asteroid ($D_{\rm a}=400$ m), $D_{\rm i, max}$ and $D_{\rm i, min}$ are 10 and 3.4 m, respectively.

\begin{figure}
\begin{center}
\includegraphics[width=\hsize]{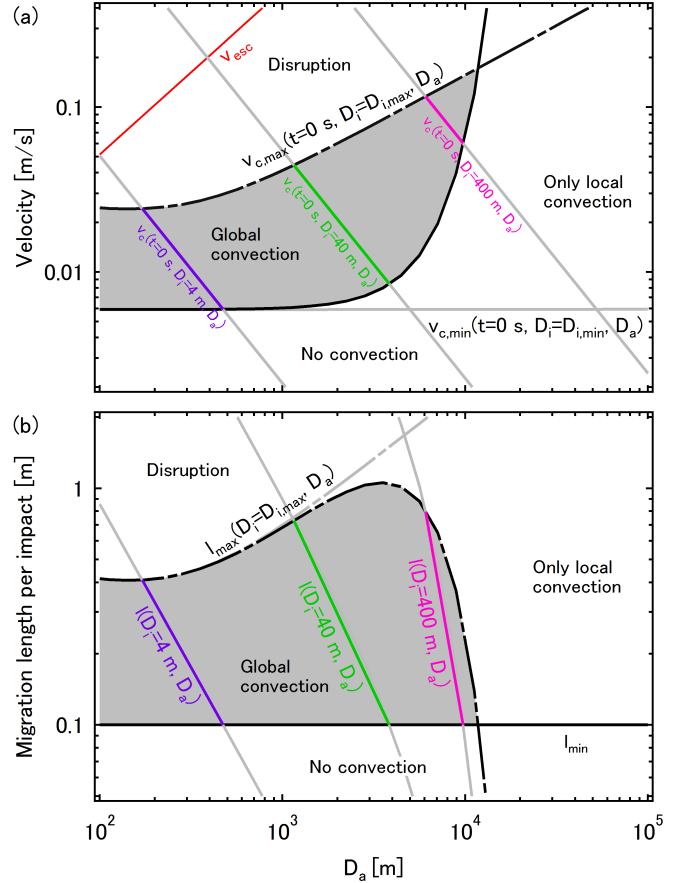}
\end{center}
\caption{(a) Regolith-convection velocity $v_{\rm c}$ on MBAs at the impact moment ($t=0$). The chain curve represents the maximum convective velocity $v_{\rm c,max}$ which can be generated by the catastrophic impact ($D_{\rm i}=D_{\rm i, max}$). The solid curve represents the minimum convective velocity $v_{\rm c,min}$ by $D_{\rm i}=D_{\rm i, min}$. The colored solid lines indicate $v_{\rm c}$ by the impactor with diameter $D_{\rm i}=$ $400$ m (pink), $40$ m (green), and $4$ m (purple) vs.~the target asteroid of diameter $D_{\rm a}$. The red solid line represents surface escape velocity $v_{\rm esc}$ on the target asteroids. The horizontal gray line shows $v_{\rm c,min}$ computed with $D_{\rm i, min}$ using the approximation $\tau_{\rm diff} = 0$. (b) Migration length per impact $l$ of MBAs. The chain curve represents the maximum of $l$ by the maximum impactor $D_{\rm i}=D_{\rm i, max}$. The horizontal solid line represents the minimum limit of migration length $l_{\rm min}$. The colored solid lines indicate $l$ by various size impactors. The gray chain curve shows $l_{\rm max}$ computed with the approximation $\tau_{\rm diff} = 0$. The parameter values used in (a) and (b) are identical to those in Fig.~\ref{fig:gamma_t}.}
\label{fig:Vc_L_Da}
\end{figure}

In Fig.~\ref{fig:Vc_L_Da}(a), the ideal convective velocities at the impact moment ($t=0$) are shown by three colored lines ($D_{\rm i}=$ $400$ m (pink), $40$ m (green), and $4$ m (purple)). The hatched areas correspond to the global seismic shaking regime. As shown in Fig.~\ref{fig:Vc_L_Da}(a), the typical velocity for small asteroids is in the order of $10^{-2}$ m/s. This convective velocity is considerably smaller than the escape velocity of the target asteroid $v_{\rm esc}=\sqrt{\frac{2}{3}\pi G \rho_{\rm a}}D_{\rm a}$ as long as $D_{\rm a}$ is greater than $10^2$ m. Thus, regolith grains would not disperse to space by regolith convection while a few grains may be released from the surface to space due to the spallation. In addition to the spallation at the very surface, surface regolith layer might experience the long-time lofting~\citep{Garcia2015}. In such a situation, the convective migration might be significantly reduced. However, the quantitative modeling of this effect is still difficult since the details of the lofting (e.g., how thick it is, how long it lasts depending on vibration conditions) have not yet been revealed well. Therefore, we neglect the lofting effect in the current model. However, its quantitative modeling is one of the most important future steps to improve the model. 

The corresponding convective migration length per impact is shown in Fig.~\ref{fig:Vc_L_Da}(b). The color and line codes used in Fig.~\ref{fig:Vc_L_Da}(b) are identical to Fig.~\ref{fig:Vc_L_Da}(a). The estimated maximum value of $l$ is about 1 m, which coincides with the assumed convective roll size. Thus, the convective resurfacing can almost be completed by a single impact event if the impactor size is sufficiently large $D_{\rm i} \simeq D_{\rm i,max}$. However, such a large scale impact rarely occurs. As discussed later in section~\ref{subsec:analysis}, the accumulation of small migrations is actually more effective than the single large impact. To compare the contributions by small and large impactors, the impact frequency $N_{\rm p}$ has to be multiplied to $l$.

\subsection{Resurfacing timescale}
\label{subsec:timescale}
Finally, by unifying all the steps mentioned above, we can obtain the resurfacing timescale as a function of the target asteroid size, $T(D_{\rm a})$. Dividing the convective roll size $A$ by the mean migration length per year $L(D_{\rm a})$, $T(D_{\rm a})$ can be calculated as, 
\begin{equation}
T(D_{\rm a})=\frac{A}{L(D_{\rm a})}.
\label{eq:timescale}
\end{equation}
Here, $L(D_{\rm a})$ can be computed by integrating the product of $l(D_{\rm i}, D_{\rm a})$ and $N_{\rm p}(D_{\rm i}, D_{\rm a})$. This procedure is described as, 
\begin{equation}
L(D_{\rm a})=\sum_{D_{\rm i}=D_{\rm i, min}}^{D_{\rm i, max}} l(D_{\rm i}, D_{\rm a})N_{\rm p}(D_{\rm i}, D_{\rm a}).
\label{eq:V_sum}
\end{equation}
The range of numerical integration in Eq.~(\ref{eq:V_sum}) spreads from $D_{\rm i, min}$ to $D_{\rm i, max}$. To compute the specific value of $L(D_{\rm a})$ by Eq.~(\ref{eq:V_sum}), Eqs.~(\ref{eq:I_frequency}), (\ref{eq:l2}), and the data shown in Fig.~\ref{fig:Nicum_Di} are used.

\section{Result and analysis}
\label{sec:result}
\subsection{Estimation of the resurfacing timescale}
\label{subsec:estimation}
\begin{figure}
\begin{center}
\includegraphics[width=\hsize]{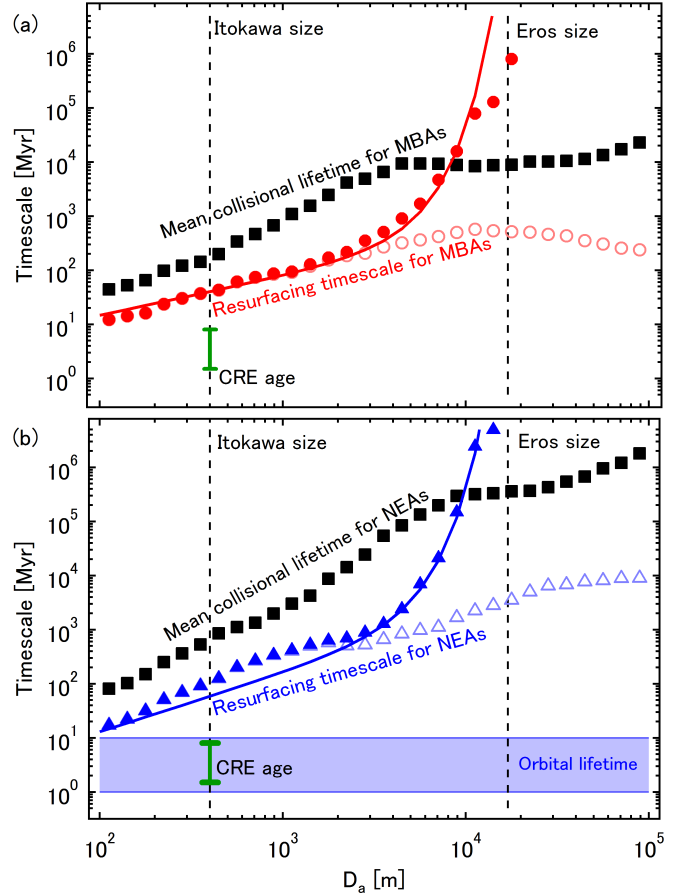}
\end{center}
\caption{(a) Various timescales for MBAs. Red solid circles represent the timescale necessary to resurface the asteroid by regolith convection, $T(D_{\rm a})$, which is numerically computed by Eq.~(\ref{eq:timescale}). Red open circles represent $T(D_{\rm a})$ by considering local convective motion as well. The red solid line shows the approximated $T(D_{\rm a})$ computed by Eq.~(\ref{eq:timescale_app_MBA}). Black solid squares represent the mean collisional lifetime for MBAs~\citep{Jutzi2010,O'Brien2005}. Population data for MBAs (red circles in Fig.~\ref{fig:Nicum_Di}) are used for the computation. An error bar at the Itokawa size indicates the cosmic-ray exposure (CRE) age of Itokawa's returned particle~\citep{Nagao2011,Meier2014}. $T(D_{\rm a})$ are computed by using the standard parameter values.
(b) Various timescales for NEAs. Blue solid triangles represent $T(D_{\rm a})$ for NEAs computed by Eq.~(\ref{eq:timescale}). Blue open triangles represent $T(D_{\rm a})$ by considering local convective motion as well. The blue solid line shows the approximated $T(D_{\rm a})$ computed by Eq.~(\ref{eq:timescale_app_NEA}). Population data for NEAs (blue triangles in Fig.~\ref{fig:Nicum_Di}) are used. $T(D_{\rm a})$ are computed by using the standard parameter values. The blue band corresponds to orbital lifetime for NEAs~\citep{Gladman2000}} 
\label{fig:timescale}
\end{figure}

The numerically computed $T(D_{\rm a})$ on the basis of OBG population data for MBAs is shown by red solid circles in Fig.~\ref{fig:timescale}(a). Similarly, $T(D_{\rm a})$ for NEAs is shown by blue solid triangles in Fig.~\ref{fig:timescale}(b). For small asteroids of $D_{\rm a}< 5 \times 10^3$ m, the resurfacing timescale approximately obeys power law since the exponential factor in Eq.~(\ref{eq:l2}) is negligible (almost unity due to the very short $\tau_{\rm diff}$). On the other hand, $T(D_{\rm a})$ tends to diverge in the large $D_{\rm a} (>5\times 10^3$ m) regime because the exponential factor in Eq.~(\ref{eq:l2}) dominates the behavior due to the large $\tau_{\rm diff}$. 

The mean collisional lifetime is also shown by black squares in Figs.~\ref{fig:timescale}(a) and (b)~\citep{Jutzi2010,O'Brien2005}. We can confirm that $T(D_{\rm a})$ necessary to resurface the asteroid is shorter than the mean collisional lifetime for both MBAs and NEAs, as long as the order of asteroid sizes is smaller than $10^4$ m. Specifically, the resurfacing timescale of Itokawa-sized asteroid $T(D_{\rm a}=400 \mbox{ m})=40$ Myr is shorter than its collisional lifetime $1.7 \times 10^{2}$ Myr. These results are insensitive to the difference between OBG and BAL populations. 

Even for the large asteroid of $D_{\rm a} > 10^4$ m in which the global seismic shaking is difficult, the resurfacing timescale might remain short if we assume the local convection can contribute to the resurfacing process. The estimated $T(D_{\rm a})$ including local convection is shown by open red circles (MBA) or open blue triangles (NEA) in Fig.~\ref{fig:timescale}. These results (open circles and triangles in Fig.~\ref{fig:timescale}) have been reported in our previous conference abstract~\citep{Yamada2015}. Actually, both estimates (with and without local convection cases) agree well in the small $D_{\rm a}$ regime since $\tau_{\rm diff}$ is sufficiently small to neglect the exponential factor in Eq.~(\ref{eq:l2}).

However, we should notice that the obtained $T(D_{\rm a})$ can vary depending on parameters that are actually very uncertain. For example, the dominant frequency $f$ might be larger than the assumed standard one particularly in small asteroids~\citep{Garcia2015}. To verify the parameter dependences, the approximated form of $T(D_{\rm a})$ will be useful. Therefore, we compute the approximated analytic form of $T(D_{\rm a})$ in the next subsection. 

\subsection{Scaling of the resurfacing timescale}
\label{subsec:analysis}
To obtain the approximated scaling form of $T(D_{\rm a})$, we begin with the power-law approximation of $N_{\rm i,cum}(\geq D_{\rm i}')$ for MBAs and NEAs as, 
\begin{equation}
N_{\rm i,cum}(\geq D_{\rm i}')=C_{\rm 1}D_{\rm i}'^{-\gamma},
\label{eq:Ni_cum_app_MBA}
\end{equation}
where the value of $C_{\rm 1}$ is a fitting parameter having dimensions of $(length)^{\gamma}$ and $\gamma$ is the characteristic exponent which corresponds to the slope of the solid line in Fig.~\ref{fig:Nicum_Di}. $C_1$ for MBAs and NEAs are obtained as $3.2\times 10^{13}$ m$^{2.8}$ and $5.2\times 10^{11}$ m$^{3.2}$, respectively. $\gamma$ for MBAs and NEAs are $2.8$ and $3.2$, respectively. Equations~(\ref{fig:Nicum_Di}), (\ref{eq:V_sum}), and (\ref{eq:Ni_cum_app_MBA}) enable us to analytically approximate $L(D_{\rm a})$ as,
\begin{eqnarray}
L(D_{\rm a}) &=& \int_{D_{\rm i, min}}^{D_{\rm i, max}} l(D_{\rm i}, D_{\rm a}) N_{\rm p}(D_{\rm i}, D_{\rm a}) \frac{dD_{\rm i}}{D_{\rm i}{\ln B}} \label{eq:V1_analy} \\
&=& \int_{D_{\rm i, min}}^{D_{\rm i, max}} l(D_{\rm i}, D_{\rm a}) P_{\rm i} \left( \frac{D_{\rm i}}{2}+\frac{D_{\rm a}}{2} \right)^{2} \nonumber \\
&& \times \left( -\frac{dN_{\rm i,cum}(\geq D_{\rm i}')}{dD_{\rm i}/D_{\rm i}{\ln B}} \right)  \frac{dD_{\rm i}}{D_{\rm i}{\ln B}}, 
\label{eq:V2_analy} 
\end{eqnarray}
where the minus sign in front of differential term of $N_{\rm i,cum}$ is for the negative slope of $N_{\rm i,cum}(\geq D_{\rm i}')$ (i.e., $\gamma > 0$), and $B=10^{0.1}\simeq 1.3$ is the constant ratio between $D'_{{\rm i}, k}$ and $D'_{{\rm i}, k+1}$. The factor $\ln B$ comes from the logarithmic binning effect, $d\log_{B}D_{\rm i}=dD_{\rm i}/D_{\rm i}\ln B$~\citep{O'Brien2005}. Substituting Eqs.~(\ref{eq:l2}) and (\ref{eq:Ni_cum_app_MBA}) into Eq.~(\ref{eq:V2_analy}), the integration can be performed analytically. 

Eq.~(\ref{eq:V2_analy}) yields twelve terms that depend on $D_{\rm i, max}$ or $D_{\rm i, min}$. By comparing the orders of these terms, a principal term dominating the behavior of $L(D_{\rm a})$ can be identified. Using the dominant term, $L(D_{\rm a})$ is approximated as,
\begin{eqnarray}
L(D_{\rm a}) &\simeq& C_{\rm \ast} Q \eta^{\alpha} f^{-1} d^{-\alpha-\beta+\frac{1}{2}} v_{\rm i}^{2\alpha} A^{\beta} G^{-\alpha+\frac{1}{2}} \nonumber \\
&& \times \rho_{\rm i}^{\alpha} \rho_{\rm a}^{-2\alpha+\frac{1}{2}} P_{\rm i} C_{\rm 1} \nonumber \\
&& \times D_{\rm a}^{-4\alpha+\frac{5}{2}} D_{\rm i, min}^{3\alpha-\gamma} {\exp}\left( -\frac{4\alpha \pi f }{Q} \tau_{\rm diff} \right),  
\label{eq:V3_analy}
\end{eqnarray}
where $C_{\ast}$ is a dimensionless constant,
\begin{equation}
C_{\ast} = -\frac{\gamma C_{\rm 0} \left( \frac{2}{3}\pi \right)^{\alpha+\frac{1}{2}}3^{2\alpha}}{4\left(3\alpha-\gamma \right) 2\alpha \left( 2\pi \right)^{2\alpha+1}}.
\label{eq:C_ast}
\end{equation}
In Eq.~(\ref{eq:V3_analy}), the dominant term depends on $D_{\rm i,min}$. This implies that the small impactors play an essential role for the regolith resurfacing process. Since $D_{\rm i,min}$ is determined by $l_{\rm min}$ through Eq.~(\ref{eq:l2}), a little more computation is necessary to obtain the final form for the analytic approximation of $L(D_{\rm a})$. Then, $T(D_{\rm a})$ can be readily calculated by Eq.~(\ref{eq:timescale}).

From the above approximation and a little more algebra, we finally obtain an approximated scaling form of $T(D_{\rm a})$ expressed by the product of simple power-law forms and an exponential form. The obtained form is written as,
\begin{eqnarray}
T(D_{\rm a}) &=& C_{\rm \ast \ast} Q^{-\frac{\gamma}{3\alpha}} \eta^{-\frac{\gamma}{3}} n^{\frac{\gamma}{3\alpha}-1} f^{\frac{\gamma}{3\alpha}} d^{\frac{\gamma}{3}+\frac{\beta\gamma}{3\alpha}-\frac{\gamma}{6\alpha}} \nonumber \\
&& \times v_{\rm i}^{-\frac{2\gamma}{3}} A^{\frac{\gamma}{3\alpha}-\frac{\beta \gamma}{3\alpha}} G^{\frac{\gamma}{3}-\frac{\gamma}{6\alpha}} \rho_{\rm i}^{-\frac{\gamma}{3}} \rho_{\rm a}^{\frac{2\gamma}{3}-\frac{\gamma}{6\alpha}} \nonumber \\
&& \times P_{\rm i}^{-1} C_{\rm 1}^{-1} D_{\rm a}^{\frac{4\gamma}{3}-\frac{\gamma}{6\alpha}-2} {\exp}\left( \frac{4\gamma \pi f \tau_{\rm diff}}{3 Q } \right),
\label{eq:timescale_app_character_MBA}
\end{eqnarray}
where $C_{\rm \ast \ast}$ is a dimensionless constant, 
\begin{equation}
C_{\ast \ast} = -\frac{4\left (3\alpha-\gamma \right)}{\gamma}\left\{ \frac{C_{\rm 0} \left( \frac{2}{3}\pi \right)^{\alpha+\frac{1}{2}}3^{2\alpha}}{2\alpha \left( 2\pi \right)^{2\alpha+1}} \right\}^{-\frac{\gamma}{3\alpha}}.
\label{eq:C_astast}
\end{equation}

Substituting $\alpha=0.47$, $\beta=0.82$ (experimentally obtained values), and $\gamma=2.8$, the specific form of Eq.~(\ref{eq:timescale_app_character_MBA}) for MBAs is obtained as,
\begin{equation}
T(D_{\rm a}) = \frac{C_{\rm \ast \ast}n^{0.97}A^{0.35}f^{1.97}d^{1.56}\rho_{\rm a}^{0.87}}{Q^{1.97}\eta^{0.93}v_{\rm i}^{1.85}\rho_{\rm i}^{0.87}P_{\rm i}G^{0.06}C_{\rm 1}} D_{\rm a}^{0.72} {\exp}\left( \frac{3.7\pi f \tau_{\rm diff}}{Q} \right),
\label{eq:timescale_app_MBA}
\end{equation}
where $C_1=3.2 \times 10^{13}$ m$^{2.8}$ is used for MBAs. The solid red curve in Fig.~\ref{fig:timescale}(a) represents Eq.~(\ref{eq:timescale_app_MBA}). Using the identical $\alpha$ and $\beta$ values, and $\gamma=3.2$, Eq.~(\ref{eq:timescale_app_character_MBA}) for NEAs is also computed as, 
\begin{equation}
T(D_{\rm a}) = \frac{C_{\rm \ast \ast}n^{1.23}A^{0.40}f^{2.23}d^{1.76}\rho_{\rm a}^{0.98}}{Q^{2.23}\eta^{1.05}v_{\rm i}^{2.10}\rho_{\rm i}^{1.05}P_{\rm i}G^{0.07}C_{\rm 1}} D_{\rm a}^{1.08} {\exp}\left( \frac{4.2\pi f \tau_{\rm diff}}{Q } \right),
\label{eq:timescale_app_NEA}
\end{equation}
where $C_1=5.2 \times 10^{11}$ m$^{3.2}$ is used for NEAs. The solid blue curve in Fig.~\ref{fig:timescale}(b) represents Eq.~(\ref{eq:timescale_app_NEA}). We can confirm that both approximated forms of $T(D_{\rm a})$ agree very well with numerically computed ones as long as we neglect the local seismic shaking effect. Moreover, Eqs.~(\ref{eq:timescale_app_MBA}) and (\ref{eq:timescale_app_NEA}) allow us to evaluate the various parameter dependences of $T(D_{\rm a})$ for MBAs and NEAs, respectively.

\section{Discussion}
\label{sec:discussion}
\subsection{Comparison with other timescales}
\label{subsec:comparison}
We think that the asteroids have been resurfaced during their stay in the main belt. Since the orbital lifetime for NEAs is ranging from $1$ Myr to $10$ Myr~\citep{Gladman2000}, $T(D_{\rm a})$ for NEAs in Fig.~\ref{fig:timescale}(b) is greater than the orbital lifetime. This means that it is difficult to resurface the asteroids within their orbital lifetime for NEAs. Thus, we basically consider that the resurfacing is important for MBAs, and discuss hereafter on MBAs values although both of Eros and Itokawa currently belong to NEAs.

In Fig.~\ref{fig:timescale}, the mean collisional lifetime (for both MBAs and NEAs) is longer than $T(D_{\rm a})$. However, the mean collisional lifetime depends on the model of $Q_{d}^{\ast}$. As mentioned in Section~\ref{subsubsec:convection}, the rubble-pile asteroid is more fragile than the monolith asteroid. Thus, $Q_{d}^{\ast}$ for rubble-pile asteroid might be much smaller than the estimate by \citet{Jutzi2010}~(e.g., \citet{Durda1998}). Such a small $Q_{d}^{\ast}$ actually results in the short collisional lifetime. Namely, the uncertainty of $Q_{d}^{\ast}$ also affects the competition between the mean collisional lifetime and $T(D_{\rm a})$. Note that, however, $Q_{d}^{\ast}$ does not directly affect the estimate of $T(D_{\rm a})$ because it is rather dominated by $D_{\rm i,min}$.

For asteroid Itokawa, \citet{Michel2009} suggested that the timescale necessary to reproduce the current crater record \citep{Hirata2009} ranges from 75 Myr to 1 Gyr. To estimate the timescale, they used the impactor population model~\citep{O'Brien2005, BottkeJr2005}, seismic shaking model~\citep{Richardson2004, Richardson2005}, and crater scaling law~\citep{Holsapple1993,Nolan1996}. The estimated timescale corresponds to the surface age computed from the number density of craters. They found that the surface age of Itokawa is equivalent to or larger than the corresponding collisional lifetime in the main belt. In addition, the resurfacing timescale of Itokawa is much shorter than this crater-based surface age. This suggests that the regolith convection may keep the surface of Itokawa fresh by the resurfacing during the crater erasure process. Moreover, the characteristic surface structure such as impact craters might be erased by the regolith convection. Then, the relatively small craters of diameter smaller than $10$ m might be erased on the surface of small asteroids because the assumed convective roll size is in the order of $10^0$ m. Indeed, the number density of craters smaller than 10 m shows an unexplained decrease on Itokawa's surface~\citep{Hirata2009, Michel2009}.

The value of resurfacing timescale on Itokawa $40$ Myr is longer than its CRE age (1.5 - 8 Myr)~\citep{Nagao2011, Meier2014}. If we assume that the regolith grains have been irradiated with cosmic rays while they are on the surface (upper part) of Itokawa in the course of convective motion, the resurfacing timescale should be equal to or longer than the CRE age. Thus, this relatively young CRE age is consistent with the current regolith convection model. 

For asteroid Eros, \citet{Richardson2004} and \citet{O'Brien2006} estimated the surface age to reproduce the observed crater record reported by~\citet{Chapman2002}. The estimated surface age for Eros is 400 $\pm$ 200 Myr~\citep{Richardson2004} or 120 Myr~\citep{O'Brien2006}. Although the difference between these two ages originates from the difference of the used crater scaling laws, both ages are much shorter than the mean collisional lifetime ($8.9\times 10^3$ Myr) in MBA. These estimated surface ages are far below the resurfacing timescale $T(D_{\rm a}=1.7\times 10^4 \mbox{ m})=6.6 \times 10^{5}$ Myr. And, this resurfacing timescale ($6.6 \times 10^{5}$ Myr) is much longer than the mean collisional lifetime as well.  This means that it is difficult to resurface the asteroid Eros by global regolith convection. However, by considering the effect of the local regolith convection in the vicinity of impact point (open circles in Fig.~\ref{fig:timescale}(a)), $T(D_{\rm a})$ for asteroid Eros ($5.2 \times 10^2$ Myr) would be close to these estimated surface ages.

\subsection{The evaluation of resurfacing feasibility}
\label{subsec:feasibility}
As already mentioned in the last paragraph of Section~\ref{subsec:estimation}, $T(D_{\rm a})$ involves a number of uncertain parameters. Thus, the effect of parameter uncertainties to the feasibility of resurfacing by regolith convection should be discussed. To evaluate the uncertainty effect, we first reconfirm the range of each parameter uncertainty in Section \ref{subsubsec:uncertainty_ranges}. Then, we introduce a dimensionless coupling parameter to simply parametrize the possibility of regolith convection (Section \ref{subsubsec:Yparameter}). Finally, the feasibility of convective resurfacing is discussed in the parameter space and the dimensionless coupling parameter (Section~\ref{subsubsec:scaled_parameter}).
\subsubsection{The ranges of parameter uncertainty}
\label{subsubsec:uncertainty_ranges}
 Eqs.~(\ref{eq:timescale_app_MBA}) and (\ref{eq:timescale_app_NEA}) allow us to read the relative parameter sensitivities in the form of $T(D_{\rm a})$ for MBAs and NEAs, respectively. The larger exponent value implies the stronger dependency in each factor of Eqs.~(\ref{eq:timescale_app_MBA}) and (\ref{eq:timescale_app_NEA}). In this sense, $Q$, $f$, $d$, and $v_{\rm i}$ are relatively sensitive parameters, whereas $\rho_{\rm a}$, $\rho_{\rm i}$, $n$, $A$, and $\eta$ are relatively insensitive parameters. 
Actually, the values of $Q$, $f$, $n$, $A$, and $\eta$ are very uncertain. Ranges of these uncertain parameters are tabulated in Table~\ref{table:parameters2}. The values of $d$, $v_{\rm i}$, $\rho_{\rm a}$ and $\rho_{\rm i}$ can somewhat be estimated by observation and/or appropriate modelings. Thus, the accurate estimate of $Q$ and $f$ is the most important factor for the reliable estimate of $T(D_{\rm a})$. For the quality factor $Q$, we consider that its order ranges from $200$ to $2000$ by following~\citet{Richardson2004} and \citet{Richardson2005}. We consider that S-type asteroids should not be dissipative due to the lack of dissipative components. Thus, $Q=2000$ would be more plausible. For the worse case estimate, we consider the case $Q=200$ as well. The dominant seismic frequency computed by~\citet{Richardson2004} and~\citet{Richardson2005} is $f=10 - 20$ Hz. However, the recent numerical study revealed that $f$ could be greater than 20 Hz (perhaps over 100 Hz) particularly for small asteroids~\citep{Garcia2015}. The value of $f$ seems to strongly depend on the internal structure of target asteroids. Therefore, further systematic investigation is necessary to estimate the reliable value of the dominant vibrational frequency. Here, we assume that $f$ might vary from $10$ to $200$ Hz. The asteroids considered in this study is highly inhomogeneous with low dissipation.

\begin{table}[hbtp]
  \caption{Lower and upper limits of parameters used in the model.}
  \label{table:parameters2}
  \centering
  \begin{tabular}{lccc}
	\hline
	 & lower limit & standard & upper limit \\
	\hline \hline
	$Q$ & 200 & 2000 & 2000\\
	$f$ (Hz) & 10 & 20 & 200 \\
	$\eta$ & $10^{-6}$ & $10^{-4}$ & $10^{-1}$ \\
	$A/d$ & \multicolumn{3}{c}{100} \\
	$n$ & 0.001 & 0.1 & 0.1 \\
	\hline
  \end{tabular}
\end{table}

The value of $\eta$ ranges from $10^{-6}$ to $10^{-1}$~\citep{Richardson2004, Richardson2005}. Because the uncertainty of $\eta$ spreads over five orders of magnitude, it may also significantly affect the estimate of $T(D_{\rm a})$. \citet{Yasui2015} have recently reported that $\eta$ ranges from $5 \times 10^{-5}$ to $5 \times 10^{-4}$. Although this result is based only on the relatively low impact velocity experiment, the estimated value is consistent with the assumed standard value. 

In addition, $A$ and $n$ are particular parameters in the present model. For the convective roll size $A$, we assume $A/d = 100$ on the basis of experimental observations~\citep{Yamada2014,Aoki1996}. However, the physical condition determining the intrinsic convective roll size has not yet been revealed. As for $n$, while we temporarily assume $n=0.1$, the reason for employing this value is not firmly based. The basis of this speculation ($n=0.1$) is the preliminary observation of the tapped granular layer~\citep{Iikawa2015}. According to the experimental result, however, very small $l_{\rm min}(\simeq 0.001 A)$ may also be able to produce a well-ordered convective structure by the accumulation of intermittent tappings. Since $T(D_{\rm a})$ is proportional to $n^{0.97}$ (Eq.~(\ref{eq:timescale_app_MBA})) or $n^{1.23}$ (Eq.~(\ref{eq:timescale_app_NEA})), the smaller $n$ results in the shorter $T(D_{\rm a})$. This tendency does not significantly affect the interpretation of the current result, viz., we employ the stricter condition for $n$ in the current estimate. However, note that the experimental observation is still quite preliminary. The accurate determination of $A$ and $n$ is a very important future problem to be investigated. 

As discussed in Section~\ref{sec:introduction}, gravity dependence of granular convection is still under debate. Although here we employ the values of $C_0$, $\alpha$, and $\beta$ obtained by~\citet{Yamada2014}, these values are also uncertain and can affect the timescale estimate~(Eqs.~(\ref{eq:timescale_app_character_MBA}) and (\ref{eq:C_astast})).  

\subsubsection{Dimensionless parameter for resurfacing difficulty}
\label{subsubsec:Yparameter}
Next, we can introduce a dimensionless coupling parameter for evaluating the feasibility of asteroid resurfacing by regolith convection. In the following calculation, we use MBAs value $\gamma=2.8$ as well as  $\alpha=0.47$ and $\beta=0.82$ to clearly observe the parameter dependences. Since the form for NEAs is basically identical to that for MBAs, the conversion to the form for NEAs is straightforward. By neglecting the exponential factor, a non-dimensionalized form of Eq.~(\ref{eq:timescale_app_MBA}) can be described as, 
\begin{equation}
T\sqrt{G\rho_{\rm a}} = Y\left(\frac{P_{\rm i}D_{\rm a}^{3}}{v_{\rm i}} \right)^{-1} (C_{1}D_{\rm a}^{-2.78})^{-1.02} \left(\frac{G\rho_{\rm a}D_{\rm a}^{2}}{v_{\rm i}^{2}}\right)^{0.44} \left(\frac{\rho_{\rm i}}{\rho_{\rm a}} \right)^{-0.93},
\label{eq:timescale_dl}
\end{equation}
where $Y$ is a dimensionless parameter which consists of principal uncertain parameters, $Q$, $\eta$, $n$, $f$, $d$, and $A$. Then, the specific form of $Y$ can be obtained as, 
\begin{equation}
Y = C_{\rm **} Q^{-1.97} \eta^{-0.93} n^{0.97} \left( \frac{f}{v_{\rm i}C_{1}^{-0.36}} \right)^{1.97} (dC_{1}^{-0.36} )^{1.56} (AC_{1}^{-0.36} )^{0.35}. 
\label{eq:Y}
\end{equation}

This dimensionless parameter $Y$ indicates the difficulty of regolith convection. Approximated $T(D_{\rm a})$ shown in Fig.~\ref{fig:timescale}(a) (the red curve) corresponds to the case of $Y \simeq 10^{-5}$. For NEAs, the blue curve in Fig.~\ref{fig:timescale}(b) corresponds to $Y \simeq 10^{-7}$. Actually, $T(D_{\rm a})$ with $Y \simeq 10^{-4}$ for MBAs (or $Y \simeq 10^{-6}$ for NEAs) becomes comparable to the mean collisional lifetime for MBAs (or NEAs). This implies that it is difficult to resurface the asteroid with $-\log_{10}Y < 4$ for MBAs (or $-\log_{10} Y < 6$ for NEAs) because such an asteroid would be disrupted by the fatal collision before its resurfacing. If $-\log_{10} Y$ is greater than $4$ for MBAs (or $6$ for NEAs), the regolith convection can be a relevant process for resurfacing the asteroids. Using the $-\log_{10} Y$ value, we can quickly evaluate the plausibility of the resurfacing by regolith convection once the accurate parameter values are obtained by explorations, experiments, numerical modelings, and so on.  

\subsubsection{Scaled parameters to resurfacing}
\label{subsubsec:scaled_parameter}
\begin{figure}
\begin{center}
\includegraphics[width=\hsize]{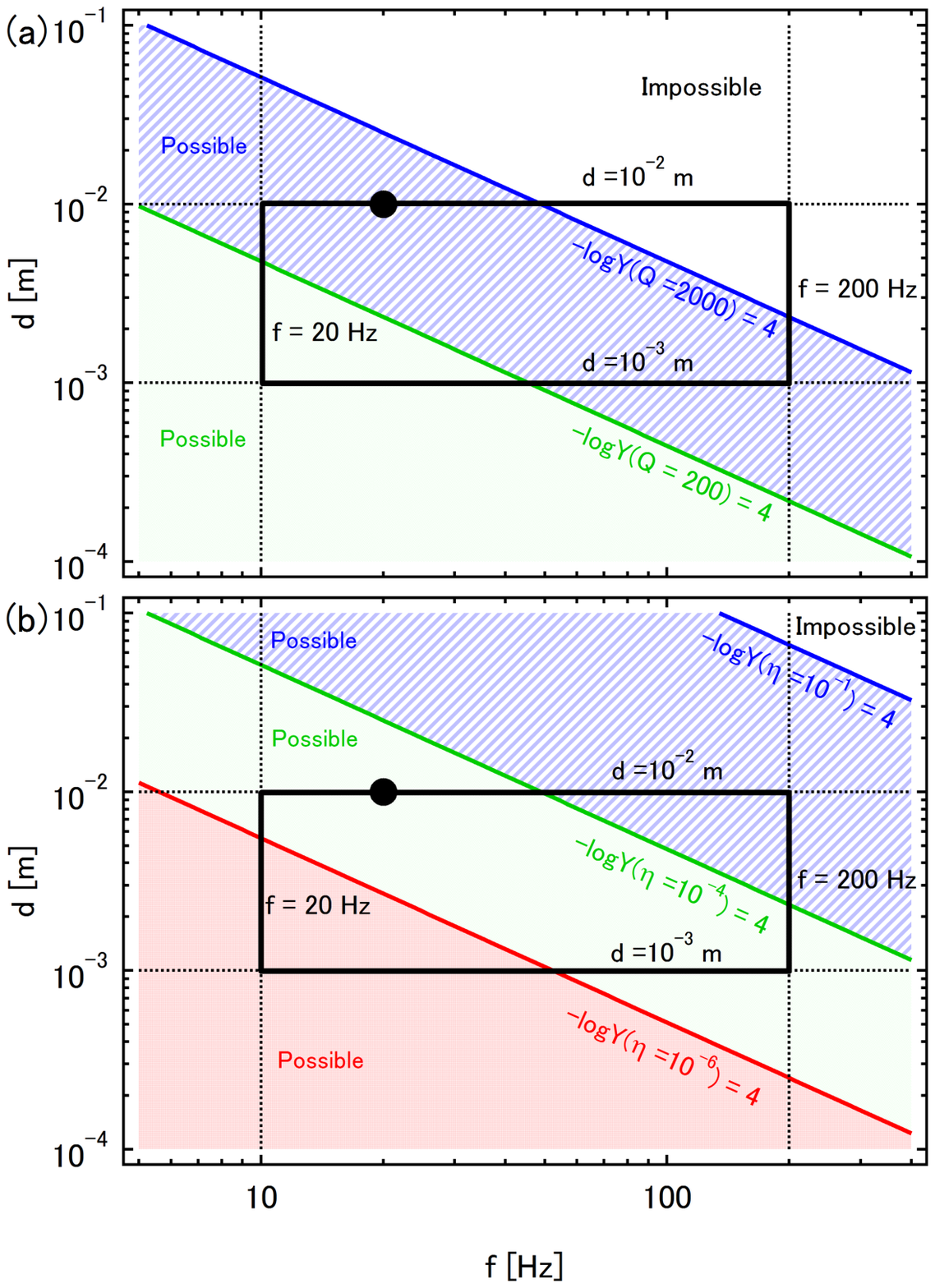}
\end{center}
\caption{Resurfacing diagram in ($d$~vs.~$f$) parameter space. The square region corresponds to the range of uncertainty of parameters ($10^{-3}\leq d \leq 10^{-2}$ m~\citep{Fujiwara2006,Yano2006} and $10 \leq f \leq 200$ Hz~\citep{Richardson2004,Richardson2005,Garcia2015}). The black solid circles represent the standard point ($d = 10^{-2}$, $f = 20$ Hz). The colored solid lines represent the resurfacing threshold level ($-\log_{10} Y = 4$) at which the resurfacing timescale becomes comparable to the mean collisional lifetime. These lines can be computed from Eq.~(\ref{eq:Y}). The hatched area corresponds to the resurfacing-possible regime ($-\log_{10} Y \geq 4$) with respective parameter, (a) $Q$ and (b) $\eta$. Other parameters are fixed to the standard ones.
} 
\label{fig:parameter_space}
\end{figure}

To systematically evaluate the possibility of asteroid resurfacing by regolith convection, we use Eq.~(\ref{eq:Y}) and the criterion $-\log_{10} Y \geq 4$ in parameter spaces. As discussed in section~\ref{subsubsec:uncertainty_ranges}, $Q$, $f$, and $d$ are very sensitive parameters to estimate the resurfacing timescale. In addition, $\eta$ has an extremely wide range of uncertainty. Thus, we focus on these parameter dependences in the following. In Fig.~\ref{fig:parameter_space}(a), the resurfacing threshold $-\log_{10} Y = 4$ is drawn in the ($d$~vs.~$f$) parameter space. The threshold levels with $Q=2000$ and $200$ are then drawn by blue and green lines, respectively. Other parameter values are fixed to the standard ones, i.e., $\eta=10^{-4}$ is used in Fig.~\ref{fig:parameter_space}(a). In the region below the threshold level, the resurfacing can take place within the mean collisional lifetime.

At the standard parameter point (solid circle in Fig.~\ref{fig:parameter_space} (a)), $-\log _{10} Y$ is sufficiently larger than 4 because the threshold level is much above this point  (blue line in Fig.~\ref{fig:parameter_space} (a)). The central square region in Fig.~\ref{fig:parameter_space}(a) ($10^{-3} \leq d \leq 10^{-2}$ mm $\times$ $10 \leq f \leq 200$ Hz) corresponds to the parameter uncertainty range considered in this study. We find that most of the square region is covered by the hatched area (regime below the threshold level with $Q=2000$). This suggests that the resurfacing by regolith convection can occur in various S-type of asteroids. Even if the dominant frequency of seismic shaking is quite high ($200$ Hz) as reported by~\citet{Garcia2015}, small regolith grains ($d \simeq 10^{-3}$ m) can be resurfaced within the mean collisional lifetime. However, there is a small regime at which the resurfacing timescale becomes longer than the mean collisional lifetime in Fig.~\ref{fig:parameter_space}(a). The white triangular region at the top right corner in the uncertain square corresponds to this hardly resurfaced regime.

Although the $\eta$ does not have a strong sensitivity in Eq.~(\ref{eq:timescale_app_MBA}), its value has an extremely wide range of uncertainty. To evaluate the effect of uncertainty, we draw another threshold levels in the identical parameter-space diagram in Fig.~\ref{fig:parameter_space}(b). In Fig.~\ref{fig:parameter_space}(b), the threshold levels ($-\log Y = 4$) are shown with $\eta = 10^{-1}$, $10^{-4}$, and $10^{-6}$. Other parameter values are fixed to the standard ones, i.e., $Q=2000$ is used in Fig.~\ref{fig:parameter_space}(b). If $\eta$ is sufficiently large, the resurfacing can easily occur within the mean collisional lifetime even if the seismic frequency is high ($200$ Hz). However, if the energy transfer is not so effective ($\eta=10^{-6}$), the resurfacing timescale becomes longer than the mean collisional lifetime.

As discussed above, it is difficult to conclude the possibility of asteroid resurfacing only from the currently available information. To accurately estimate the possibility, we have to restrict the parameter values in much smaller regime. In this study, we provide a simple framework by which the possibility of asteroid resurfacing by regolith convection can be evaluated once these parameter values were well determined. 
A stochastic modeling such as Monte Carlo simulation might be helpful to evaluate the possibility. The stochastic modeling could also be useful to discuss the random accumulation of the regolith migration composing a convective motion. These types of stochastic modelings are important future problems to refine the estimate reliability.

\subsection{Effects not considered in the model}
\label{subsec:intermittent}
In the model, convective motion of regolith grains consists of the accumulation of intermittent migrations. However, the current form of granular convective velocity is obtained on the basis of steady vibrated granular experiment~\citep{Yamada2014}. The granular convective velocity scaling under the truly intermittent attenuating vibration should be revealed to improve the model. The values of $C_{\rm 0}$, $\alpha$, and $\beta$ might be affected by such experimental conditions. 

There are some other issues that might significantly affect the estimate of $T(D_{\rm a})$. For instance, the effect of grains cohesion might be essential in the regolith convection under the microgravity condition~\citep{Scheeres2010}, although we neglect its effect in the current model. The effects of polydispersity and irregular shape of regolith grains are also omitted in this study. Some other factors that should be considered regarding experimental conditions such as container wall and interstitial air were also discussed in the previous paper~\citep{Yamada2014}. In~\citet{Garcia2015} or~\citet{Scheeres2010}, lofting of the regolith layer by impact-induced vibration was studied. During the lofting, regolith grains cannot make a contact network. Then, the regolith convection might hardly occur in very shallow part of the regolith layer. By the lofting effect, the convective regime (hatched areas in Figs.~\ref{fig:tv_Da},~\ref{fig:Di_range}, and~\ref{fig:Vc_L_Da}) could become smaller than the current estimate. More accurate and quantitative modeling of the lofting is a possible future work. The current model only discusses the considerably averaged picture of the global regolith convection.

\section{Conclusions}
\label{sec:conclusions}
We developed a simple model of regolith convection which enables us to estimate the resurfacing timescale caused by regolith convection. The model consists of three steps: impact, vibration, and the accumulation of intermittent convections. Although we assume the extremely averaged picture of the asteroid covered with regolith, we can estimate the resurfacing timescale of the various-sized S-type asteroids of diameter $D_{\rm a}$, using the model. To estimate the resurfacing timescale, the constant standard values are used although some of them depend on asteroid properties. In addition, we present a dimensionless parameter $Y$ that indicates the difficulty of resurfacing process. Using $Y$, the rough estimate of the possibility of resurfacing can be performed. Because a number of uncertain parameters are used in the present model, the approximate scaling form of $T$ and $Y$ are helpful to evaluate the relative importance of the parameters. As a result, we find that it is possible to resurface the asteroids by regolith convection within their mean collisional lifetime if we employ reasonable parameter value ranges. Thus, the regolith convection should be taken into account as one of the plausible resurfacing processes in the history of asteroid surface evolution. The knowledges obtained by further observations and experiments will refine the regolith convection model. In particular, the observation by {\it Hayabusa2} and {\it OSIRIS-REx} or their returned samples will bring us various helpful information to estimate much more reliable resurfacing timescale.

\section*{Acknowledgments}
We are grateful to Dr. D. P. O'Brien and Dr. W. F. Bottke for providing us the data of impact population in MBA and NEA. We would like to thank S. Watanabe, H. Kumagai, and S. Sirono for fruitful discussions and suggestions. This research has been partly supported by JSPS KAKENHI No.~15H03707, and Nagoya University Program for Leading Graduate Schools (Leadership Development Program for Space Exploration and Research).









\end{document}